\documentclass[preprint,12pt]{elsarticle}

\usepackage{epsfig}
\usepackage{amssymb}
\journal{Journal of Magnetism and Magnetic Materials}

\begin{document}
\newcommand{\mub}{$\mu_{\rm B}$}
\begin{frontmatter}

%% Title, authors and addresses

%% use the tnoteref command within \title for footnotes;
%% use the tnotetext command for theassociated footnote;
%% use the fnref command within \author or \address for footnotes;
%% use the fntext command for theassociated footnote;
%% use the corref command within \author for corresponding author footnotes;
%% use the cortext command for theassociated footnote;
%% use the ead command for the email address,
%% and the form \ead[url] for the home page:
%% \title{Title\tnoteref{label1}}
%% \tnotetext[label1]{}
%% \author{Name\corref{cor1}\fnref{label2}}
%% \ead{email address}
%% \ead[url]{home page}
%% \fntext[label2]{}
%% \cortext[cor1]{}
%% \address{Address\fnref{label3}}
%% \fntext[label3]{}

\title{The modulated antiferromagnetic structures in multiferroic FeVO$_4$: a $^{57}$Fe M\"ossbauer spectroscopy investigation}

%% use optional labels to link authors explicitly to addresses:
%% \author[label1,label2]{}
%% \address[label1]{}
%% \address[label2]{}

\author{D. Colson, A. Forget}
\author{P. Bonville} 
\ead{pierre.bonville@cea.fr}

\address{CEA, Centre de Saclay, DSM/IRAMIS/Service de Physique de l'Etat Condens\'e \\ 91191 Gif-sur-Yvette Cedex, France}

\begin{abstract}
We present a $^{57}$Fe M\"ossbauer spectroscopy study of the two incommensurate magnetic phases in the multiferroic material FeVO$_4$. We devise lineshapes appropriate for planar elliptical and collinear modulated magnetic structures and show that they reproduce very well the M\"ossbauer spectra in FeVO$_4$, in full qualitative agreement with a previous neutron diffraction study. Quantitatively, our spectra provide precise determinations of the characteristics of the elliptical and modulated structures which are in good agreement with the neutron diffraction results. We find that the hyperfine field elliptical modulation persists as $T \to$\,0, which we attribute to an anisotropy of the hyperfine interaction since a moment modulation is forbidden at $T=0$ for a spin only ion like Fe$^{3+}$.

\end{abstract}

\begin{keyword}
%% keywords here, in the form: keyword \sep keyword
multiferroics \sep M\"ossbauer spectroscopy \sep modulated magnetic structure \sep FeVO$_4$ 
%% PACS codes here, in the form: \PACS code \sep code
\PACS 75.85.+t \sep 77.55.Nv \sep 76.80.+y \sep 75.25.-j
%% MSC codes here, in the form: \MSC code \sep code
%% or \MSC[2008] code \sep code (2000 is the default)

\end{keyword}

\end{frontmatter}

%% \linenumbers

%% main text
\section{Introduction}
It is now generally accepted that most multiferroic materials, i.e. materials where magnetic and electric dipole moments are long range ordered and coupled \cite{smolenski}, are associated with non-collinear spin density waves (SDW) incommensurate with the lattice \cite{mostovoy,tokura}, like cycloidal or spiral arrangements. The weak coupling case is illustrated by BiFeO$_3$, where ferroelectric order \cite{smith} ($T_{\rm c}$ = 1143\,K) takes place at a much higher temperature than the antiferromagnetic (AF) order ($T_{\rm N}$ = 643\,K), which consists in an incommensurate cycloidal moment arrangement \cite{sosnowska,kadomtseva}. More recently, a new class of multiferroics has been discovered, pertaining to the strong coupling case where ferroelectricity is induced by the non-collinear SDW \cite{kimura,goto} and appears therefore simultaneously with the SDW order. Examples of this class are TbMnO$_3$ \cite{kimura} and TbMn$_2$O$_5$ \cite{hur}. The link between ferroelectricity and non-collinear magnetic order can be obtained in a continuum theory by considering the so-called Lifshitz invariant coupling the electric polarisation {\bf P} and the gradient $\nabla {\bf M}$ of the inhomogeneous magnetisation \cite{mostovoy}. The spontaneous polarisation can be viewed as due to an equivalent polarising electric field {\bf E}$_{LI}$ = $\gamma [({\bf M.\nabla})\ {\bf M} - {\bf M} ({\bf \nabla.M})]$, where $\gamma$ is the coupling parameter. The volume averaged polarisation can then be shown to be non-zero for spiral (elliptic) structures and to vanish for collinear SDW. 
 
Helical order usually appears in AF materials as a result of exchange frustration, when for instance first and second neighbour exchange interactions are of the same magnitude \cite{yoshimori}. Therefore, the strong coupling in AF multiferroics can be expected to be magnetically frustrated, which is reflected in the fact that the actual ordering temperature ($T_{\rm N}$) is much lower than the exchange coupling, whose magnitude is the paramagnetic Curie temperature $\vert \theta_p \vert$. For ferromagnetic interactions, helical structures are induced by antisymmetric (or Dzyaloshinski-Moriya) exchange \cite{dzialo}. Neutron diffraction is by far the best method allowing observation and characterisation of these incommensurate magnetic structures, but M\"ossbauer spectroscopy, although being a local technique which does not give access to the propagation vector, can be rather selective through lineshape analysis, especially in the case of amplitude modulated structures.

We report here on a detailed $^{57}$Fe M\"ossbauer spectroscopy study of the strong coupling multiferroic FeVO$_4$, which presents the property of showing two magnetic transitions towards incommensurate phases \cite{daoud,kundys} like TbMnO$_3$ \cite{kenzel}. The high temperature magnetic phase (15.7\,K $<$ T $<$ 23\,K, phase I) is a collinear sine-wave modulated structure which is not ferroelectric. The low temperature phase ($T <$ 15.7\,K, phase II) is a planar non-collinear elliptical structure showing a spontaneous electric polarisation. The major elliptical axis in phase II coincides with the moment direction in phase I \cite{daoud}. Early M\"ossbauer spectra have been reported in this compound \cite{levinson, robertson}, but they could not be thoroughly interpreted due to the lack of knowledge of the magnetic structure. We show here that the peculiar shapes of the M\"ossbauer spectra in both magnetic phases are entirely compatible with the magnetic structures determined by neutron diffraction \cite{daoud}. We also present the thermal variation of the characteristics of the magnetic structures. 

\section{Sample synthesis and magnetic characterisation}

The polycrystalline FeVO$_4$ sample was synthesized by heating a 1:1 molar mixture of V$_2$O$_5$ and Fe$_2$O$_3$ (hematite) at 550, 625, 700 and 715$^\circ$ during 10h at each temperature and with intermediate grindings. FeVO$_4$ crystallises in the $P\bar 1$ space group and the triclinic unit cell contains 3 different crystallographic Fe sites with very low point symmetry (inversion $\bar 1$) \cite{robertson}. All the diffraction peaks of the XRD pattern can be indexed based on the ICCD card of FeVO$_4$ (\#00-038-1372) with no trace of impurity phases. 

The magnetic susceptibility $\chi$ of FeVO$_4$ was measured in a field of 0.1\,T between 2 and 25\,K, and with a field of 1\,T from 25\,K up to room temperature, using a Cryogenic Vibrating Sample Magnetometer. 
\begin{figure}[ht]
\begin{center}
\includegraphics[width=9cm]{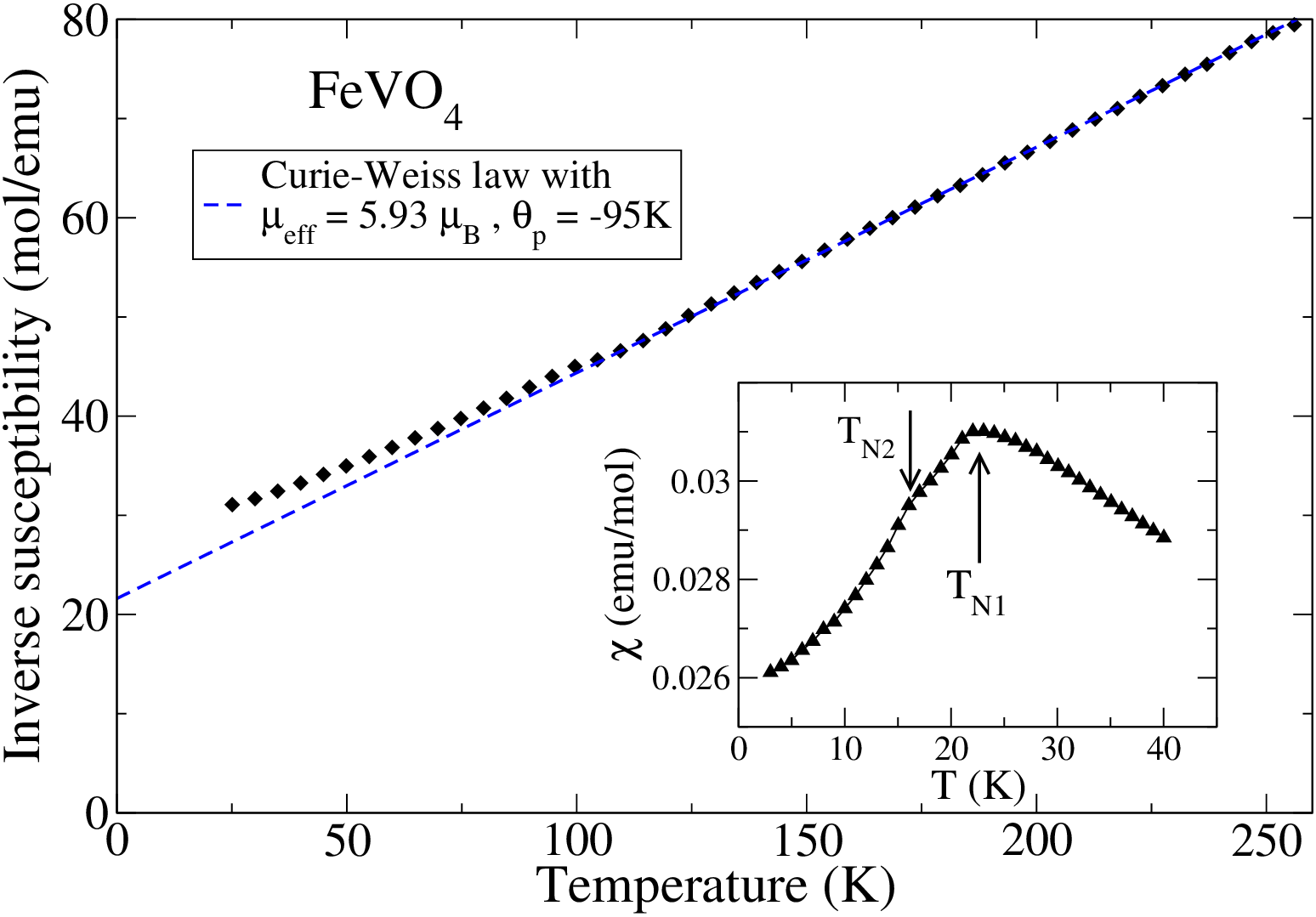}
\caption{Thermal variation of the inverse susceptibility in FeVO$_4$; the insert shows the low temperature variation of the susceptibility.}
\label{xi}
\end{center}
\end{figure}
Down to about 100\,K, the inverse susceptibility (Fig.\ref{xi}) follows a Curie-Weiss law $\chi = \frac{\mu_{eff}^2}{3k_{\rm B}(T-\theta_p)}$ with $\mu_{eff}$=5.93\,\mub\ (very close to the  effective moment 5.916\,\mub\ expected for the S=5/2 ion Fe$^{3+}$) and $\theta_p \simeq -$95\,K, characteristic of antiferromagnetic interactions. On further lowering the temperature, an anomaly occurs at $T_{\rm N1} \simeq$23\,K, marking the onset of the first transition, and an inflexion point is observed at $T_{\rm N2}$ = 15.7\,K, signalling the second transition. A remarkable feature is that the ratio $\vert \theta_p \vert/T_{\rm N}$ is close to 5, which indicates a rather high degree of frustration of the magnetic interactions, as already observed in Refs.\cite{daoud,zhang}. The deviation of 1/$\chi$ from the Curie-Weiss law below 100\,K, which cannot be attributed to crystal electric field effects absent for Fe$^{3+}$, is in line with this picture and show the persistence of strong short range spin correlations far in the paramagnetic phase \cite{zhang}.  

Before describing the M\"ossbauer data, in the two next sections we first recall the effect on the spectra of equal moment structures, then we compute the unusual lineshapes associated with incommensurate modulated magnetic structures, either planar elliptical or collinear.

\section{Spectral effects for an incommensurate equal moment spiral or cycloidal arrangement}

For the L=0, S=5/2 Fe$^{3+}$ ion in the magnetically ordered phase of insulators, the magnetic hyperfine field $H_{hf}$ at the $^{57}$Fe nucleus site is proportional to the spontaneous moment with a very good approximation, with a hyperfine constant $C_{hf} \simeq$11\,T/\mub. In the following, we shall refer equivalently to the hyperfine field or to the spontaneous moment (except at the lowest temperature, see section \ref{disc}). In the magnetically ordered phase, the M\"ossbauer spectrum associated with a static hyperfine field is a six-line pattern. The quadrupolar hyperfine interaction for $^{57}$Fe$^{3+}$ is in general much smaller than the magnetic hyperfine interaction. At first perturbation order, it gives rise to a small lineshift for each line (assuming axial symmetry for the Fe site): 
\begin{equation}
\delta e_j = \varepsilon_j \ \frac{3\Delta E_Q}{4} \ (\cos^2\theta - \frac{1}{3}).
\label{shift}
\end{equation}
In this expression, $j$ is the line index, $\varepsilon_j$ is $+$1 for the 2 external lines and $-$1 for the 4 inner lines, $\Delta E_Q$ is the quadrupole parameter whose absolute value can be measured in the paramagnetic phase and $\theta$ is the angle between the hyperfine field (or the spontaneous magnetic moment) and the principal axis of the electric field gradient (EFG) tensor at the Fe site. At second perturbation order, the lineshift is:
\begin{equation}
\delta^2 e_j= \varepsilon_j \ \frac{3 \Delta E_Q}{4} \ (1+\beta_j \frac{\Delta E_Q}{h} \ \sin^2 \theta) \cos^2 \theta,
\end{equation}
where $\beta_j$ is a coefficient depending on the specific line. The magnetic hyperfine interaction in the excited 14.4\,keV nuclear state alone enters here, through the quantity $h = \frac{1}{2} \ g_n \mu_n H_{hf}$, where $\mu_n$ is the nuclear Bohr magneton and $g_n=-0.10$ is the gyromagnetic ratio of the excited state. In the case of equal moment helical or conical structures, there is no distribution of hyperfine field values and a small spectral effect can arise from the distribution of $\theta$ values (if any) associated with the incommensurate structure. It is clear that the first order lineshift yields equivalent broadenings for all the lines. By contrast, the second order shift is different for each line through its dependence on $\beta_j$, and its spectral effect consists in inhomogeneous line broadenings. These have been observed in BiFeO$_3$ \cite{palewicz,lebeugle}, although in this case it can be shown that they are due to the anisotropy of the hyperfine interaction itself \cite{zalessky} and not to the distribution of $\theta$ values associated with the cycloidal spin structure \cite{lebeugle}. Inhomogeneous line broadenings due to a helical incommensurate magnetic structure were observed in the langasite compound Ba$_3$NbFe$_3$Si$_2$O$_{14}$ \cite{marty} and in MnGe (doped with Fe) \cite{deutsch}.

Much more spectacular effects on the lineshape arise from moment modulated structures, since then the main spectral effect is due to the distribution of hyperfine field values. This is described in the following, where the small quadrupolar line-shifts have not been considered. 

\section{M\"ossbauer lineshapes associated with incommensurate elliptical and sine wave structures}

A planar non-collinear elliptical magnetic structure is characterised by the values of the two axes of the ellipse, or by the value of the major axis $H_{hf}^{max}$ and the ratio $y=H_{hf}^{min}/H_{hf}^{max}$. Using the unit vectors {\bf a} and {\bf b} along the principal axes of the ellipse as basis vectors, the hyperfine field writes:
\begin{equation}
 H_{hf}(\theta) = H_{hf}^{max}\ (\cos\varphi\ {\bf a} +\ y\  \sin\varphi \ {\bf b}),
\label{ell}
\end{equation}
and, in the case of an incommensurate propagation vector, the angle $\varphi$ is uniformly distributed between 0 and 2$\pi$. Then, the hyperfine field distribution at the nucleus site in the interval $H_{hf}^{min} \le H_{hf} \le H_{hf}^{max}$ is given by, using $h=H_{hf}/H_{hf}^{max}$:
\begin{equation}
P_{ell}(H_{hf}) \propto \frac{1}{\vert \frac{dH_{hf}(\varphi)}{d\varphi} \vert}= \frac{h}{\sqrt{(h^2-y^2)(1-h^2)}}.
\end{equation}
\vspace{0.5cm}
\begin{figure}[ht]
\begin{center}
\includegraphics[width=12cm]{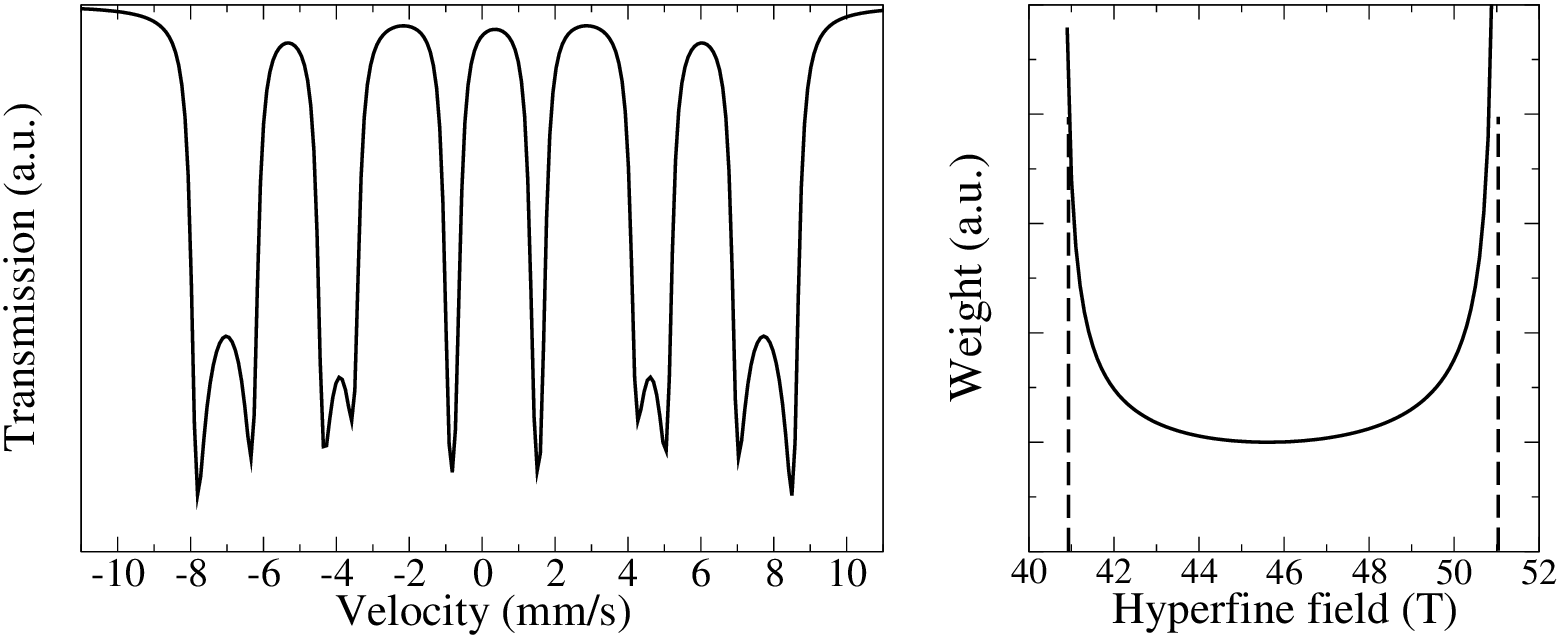}
\vspace{0.01cm}
\caption{{\bf Right panel:} Distribution of hyperfine fields associated with a planar elliptical arrangement of Fe moments with incommensurate propagation vector; the major axis is taken to be 51\,T and the ratio of minor to major axes is 0.8; {\bf Left panel}: Calculated magnetic hyperfine spectrum for such an elliptical structure.} 
\label{sp-ell}
\end{center}
\end{figure}
\begin{figure}[ht!]
\begin{center}
\includegraphics[width=12cm]{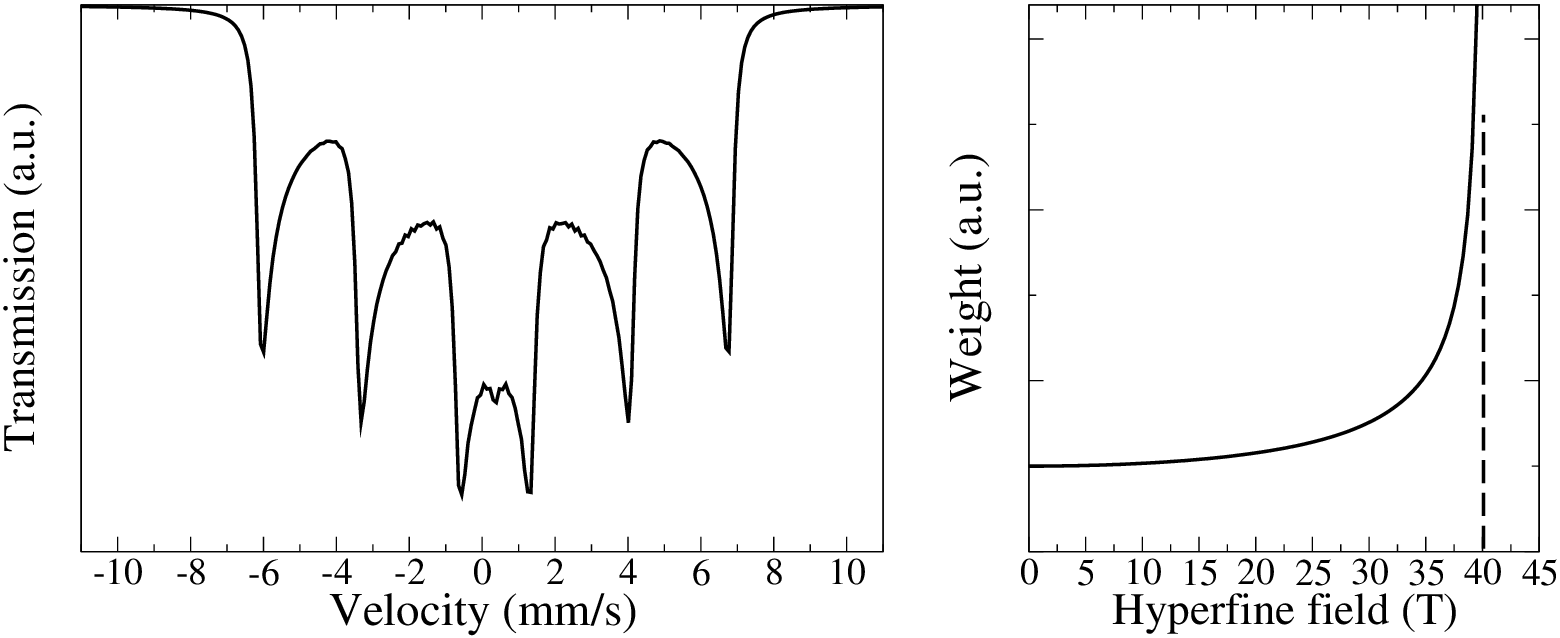}
\vspace{0.01cm}
\caption{{\bf Right panel:} Distribution of hyperfine fields associated with a collinear arrangement of Fe moments with incommensurate sine wave modulation; the maximum field is taken to be 40\,T; {\bf Left panel}: Calculated magnetic hyperfine spectrum for such a sine wave structure.}
\label{sp-sin}
\end{center}
\end{figure}
This distribution is represented on the right panel of Fig.\ref{sp-ell}, and it can be seen that it diverges at the values $H_{hf}^{min}$ and $H_{hf}^{max}$; the corresponding hyperfine spectrum, shown on the left panel, presents six lines, each of which is split in two peaks, in agreement with the shape of the distribution. 

For the case of a collinear sinusoidally modulated structure, the hyperfine field is also given by equation (\ref{ell}), but with $y$=0. It is then straightforward to see that the distribution function is given by:
\begin{equation}
 P_{sin}(H_{hf}) \propto \frac{1}{\sqrt{1-h^2}}.
\label{sin}
\end{equation}
This function is represented on the right panel of Fig.\ref{sp-sin}: it is characterised by a divergence for $H_{hf}=H_{hf}^{max}$ and by a non vanishing weight extending to $H_{hf}$=0. This explains the shape of the corresponding hyperfine spectrum shown in the left panel of Fig.\ref{sp-sin}, with a large spectral weight at zero velocity.
%\vspace{0.3cm}

\section{M\"ossbauer spectra in FeVO$_4$}

\begin{figure}[ht!]
\begin{center}
\includegraphics[width=9.cm]{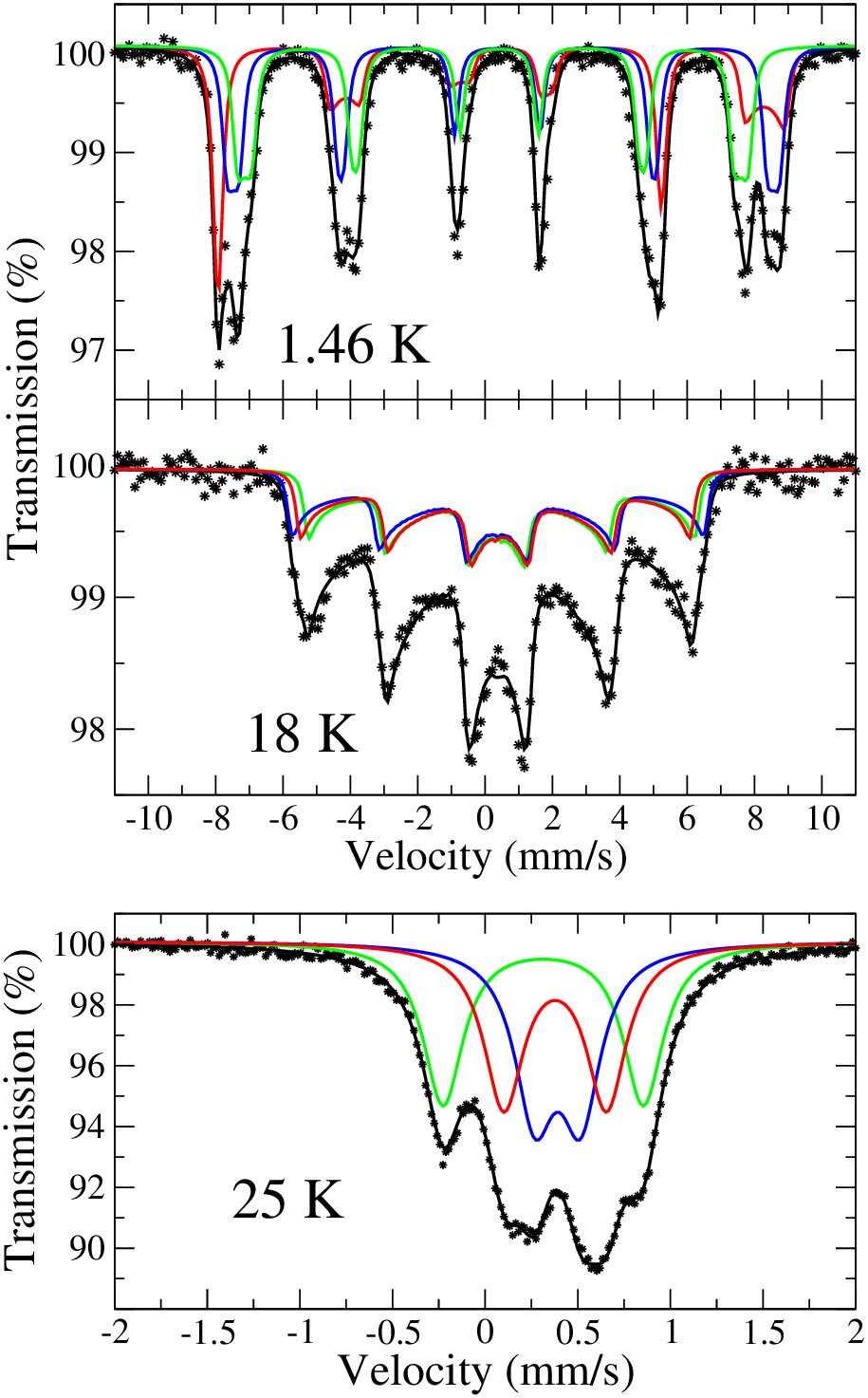}
\vspace{0.5cm}
\caption{$^{57}$Fe M\"ossbauer absorption spectra at 1.46\,K (phase II), 18\,K (phase I) and 25\,K (paramagnetic phase) in FeVO$_4$. Note the difference in velocity scales between the 25\,K spectrum and the low temperature spectra. The lines are fits with 3 subspectra to a planar elliptical magnetic structure (1.46\,K), to a collinear sine-wave magnetic structure (18\,K) and to a hyperfine quadrupole interaction (25\,K).}
\label{moss}
\end{center}
\end{figure}
Absorption M\"ossbauer spectra on the isotope $^{57}$Fe have been recorded betwen 1.46 and 25\,K, using a commercial Co$^*$:Rh $\gamma$-ray source mounted on a constant acceleration electromagnetic drive. Representative spectra in magnetic phases II (1.46\,K) and I (18\,K), and in the paramagnetic phase (25\,K) are shown in Fig.\ref{moss}. Our spectra are in good agreement with those published previously \cite{levinson,robertson}.

At 25\,K, in the paramagnetic phase, the hyperfine quadrupolar interaction alone is present, and the fit must be done with 3 equal weight subspectra corresponding to the 3 crystallographic sites of Fe in FeVO$_4$. The isomer shifts of these subspectra are around 0.46(4)\,mm/s with respect to $\alpha$-Fe, typical for trivalent Fe in insulators. The quadrupole splittings are close to those given in Ref.\cite{robertson}, namely in decreasing order: 1.08(2), 0.55(2) and 0.24(2)\,mm/s. In terms of the components $V_{ii}$, $i=X,Y,Z$, of the EFG tensor at each site, each quadrupole splitting is worth: $\vert \Delta E_Q \vert = \vert \frac{eQV_{ZZ}}{2} \vert \ \sqrt{1+\eta^2/3}$, where the asymmetry parameter is: $\eta=\vert \frac{V_{YY}-V_{XX}}{V_{ZZ}} \vert$. Due to the low site symmetry, $\eta$ is expected to be non-zero, but it is not possible to obtain independently $V_{ZZ}$ and $\eta$. In the following, we will assume $\eta$=0. In addition, the sign of $V_{ZZ}$ cannot be determined from these zero field paramagnetic phase spectra.

In the magnetic phases, the magnetic hyperfine spectra are quite different in phase I and in phase II. They are both six-line patterns, as expected for the magnetic hyperfine interaction of $^{57}$Fe, but with peculiar shapes: the spectrum in phase II (1.46\,K) seems rather complex, with its rightmost peak well resolved, and the spectrum in phase I (18\,K) bears a strong resemblance with that shown in Fig.\ref{sp-sin}. 

Therefore, we fitted the spectra in phase II to 3 equal weight subspectra associated with an incommensurate planar elliptical structure, like that shown in Fig.\ref{sp-ell}. Since the spectra are somewhat asymmetric with respect to zero velocity, quadrupolar effects should be considered for completeness. For this purpose, the knowledge of the electric field gradient (EFG) tensor at the Fe site is in principle required, but the very low symmetry at the Fe sites precludes any {\it a priori} determination. Since the quadrupolar effects can be considered as a perturbation with respect to the magnetic hyperfine interaction, we used the line shifts given by expression (\ref{shift}) for fitting of the spectra. In a frame where the z-axis is normal to the plane of the ellipse, the principal axis OZ of the EFG tensor is determined by its polar and azymutal angles $\Theta$ and $\Phi$. For a given value of the orientation $\varphi$ of the hyperfine field in the plane of the ellipse, the angle $\theta$ between the hyperfine field and OZ is such that:
\begin{equation}
\cos^2 \theta(\varphi) = \sin^2 \Theta \ \frac{\cos^2(\varphi-\Phi)}{\cos^2\varphi + y^2 \sin^2 \varphi},
\end{equation}
where $y$ is the ratio of the minor to major axis of the ellipse. For a homogeneous distribution of $\varphi$, the angle $\Phi$ results mainly in a dephasing of the $\cos^2 \theta$ values and thus has little influence on the spectrum. The fits were performed by letting $\Theta$, $\Delta E_Q$, $H_{hf}^{max}$ and $y$ as free parameters for each subspectrum. We find that the obtained quadrupolar parameter values, namely 1.4(3), 0.4(1) and $-$0.3(1)\,mm/s are not far from those measured in the paramagnetic phase. However, one must keep in mind the possibility of a lattice distortion occurring at $T_{\rm N2}$, where the ferroelectric order sets in, which could alter the quadrupolar parameter values. As to the spectra in phase I, they are correctly fitted to 3 equal weight subspectra associated with an incommensurate collinear sine wave structure like that shown in Fig.\ref{sp-sin}. In this case, the large distribution of hyperfine field values washes out the effects of the small quadrupolar interaction, which is reflected in the symmetry of the spectra with respect to zero velocity.

\begin{table}[ht]
\begin{center}
\begin{tabular}[ht]{|c||c|c|c|c|c|} \hline
        &  $H_{hf}^{max}$(T) & $y$ &  $m_A$ (\mub)  &  $y_n$ & $H_{hf}^{max}/m_A$ (T/\mub) \\ \hline \hline
site 1,II  &      52.4(1)      &  0.92(1)  &  4.51(7)    &  0.81(3)    &   11.6             \\ \hline
site 2,II  &      51.0(1)      &  0.94(1)  &  4.29(7)   &   0.79(3)    &   11.9             \\ \hline
site 3,II  &      47.0(1)      &  0.94(1)  &  4.18(6)   &   0.76(3)    &   11.3             \\ \hline\hline
site 1, I  &      38.2(1)      &        &  3.23(5)   &           &   11.8             \\
\hline
site 2, I  &      36.6(1)      &        &  3.00(5)   &           &   12.1             \\
\hline
site 3, I  &      35.5(1)      &        &  2.86(3)   &           &   12.4             \\
\hline
\end{tabular}
\end{center}
\caption{For the 3 sites of Fe in FeVO$_4$: in phase II (planar elliptical structure): maximum hyperfine field $H_{hf}^{max}$ and ratio $y = H_{hf}^{min} / H_{hf}^{max}$ at 1.46\,K from the present work, major axis of the elliptical structure $m_A$ and ratio of minor to major moment axes $y_n$ at 2\,K according to Ref.\cite{daoud}, deduced hyperfine constant; in phase I (sine wave structure) at 18\,K: maximum hyperfine field of the modulation, maximum magnetic moment from Ref.\cite{daoud}, deduced hyperfine constant.}
\label{tab1}
\end{table} 
\begin{figure}[ht!]
\begin{center}
\includegraphics[width=9.cm]{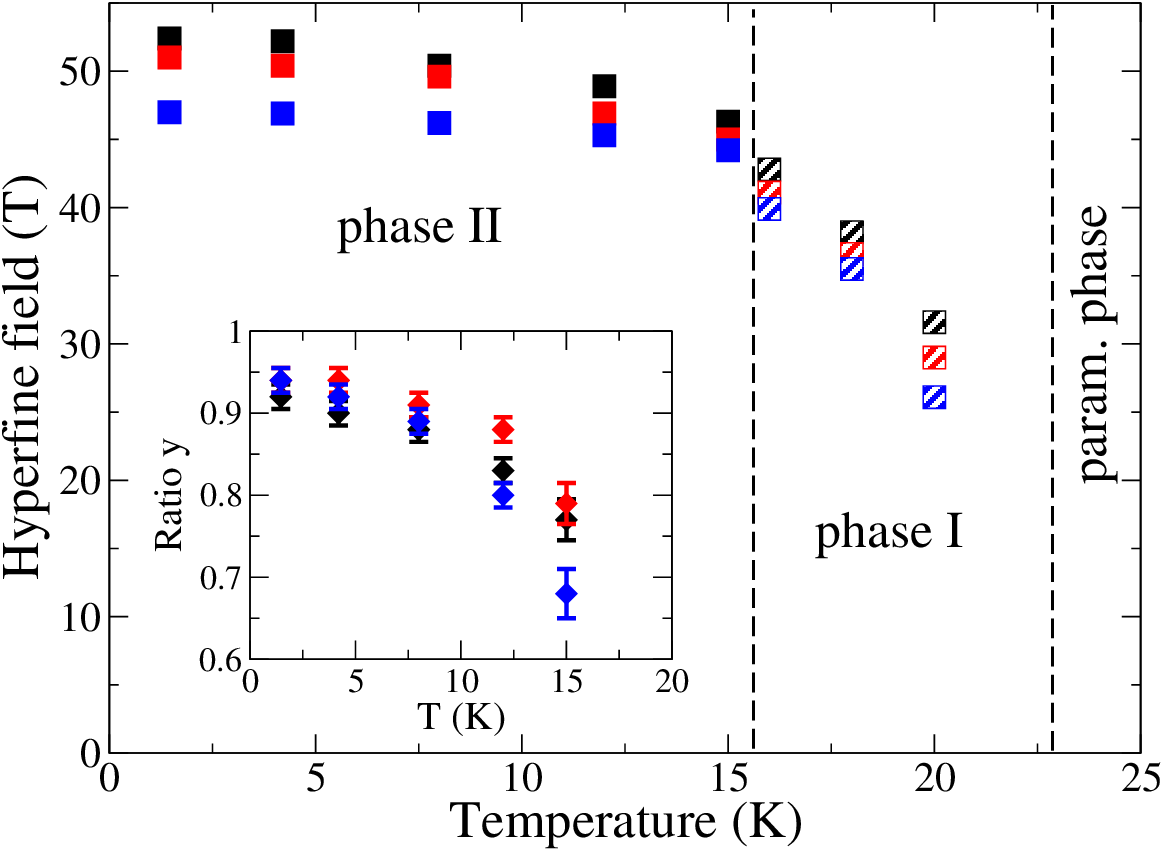}
\vspace{0.5cm}
\caption{Thermal variation of the major axis of the hyperfine field ellipse in phase II of FeVO$_4$ (full symbols) and of the maximum hyperfine field in phase I (striped symbols) for the 3 Fe sites. Due to the good statistics of the spectra, the error bar is of the size of the point. The insert shows the thermal variation of the ratio $y$ of the minor to major axes of the hyperfine field ellipse in phase II.}
\label{hhf}
\end{center}
\end{figure}
Figure \ref{hhf} shows the thermal variation of $H_{hf}^{max}$ and $y$, and Table \ref{tab1} displays the values obtained at 1.46\,K together with a comparison with those derived from neutron diffraction \cite{daoud} at 2\,K. At these base temperatures, our hyperfine field values for the major ellipse axis are in good agreement with those of the major moment axis using a hyperfine constant value $\simeq$ 11.6\,T/\mub\ close to the standard value. However, we find that the ratio of minor to major axis is close to 0.93(1) for all 3 sites, larger than the value 0.78(3) obtained by neutron diffraction. As temperature increases and the transition is approached, the ratio $y$ decreases, i.e. the ellipse is getting more and more oblate. This matches well the neutron diffraction finding that the phase II ellipse ``merges'' into the phase I collinear sine-wave, the major axis of the former becoming the moment direction of the latter. There is good continuity between phase II and phase I, and in phase I at 18\,K, the agreement with the neutron values is rather good, with a slightly higher mean hyperfine constant of 12.1\,T/\mub.

We note that our range of Fe$^{3+}$ moment values at 1.46\,K, as well as that in Ref.\cite{daoud} at 2\,K, are at odds with the upper bound of the moment value of 1.95\,\mub\ derived in the zero field $^{51}$V NMR study of Ref.\cite{zhang} in single crystal FeVO$_4$. This discrepancy could be due to an incorrect estimation of the transferred hyperfine constant at the $^{51}$V site in the ordered phase in zero field, which is taken in Ref.\cite{zhang} to be equal to the high field value in the paramagnetic phase. 

\section{Discussion} \label{disc}

For a spin only ion like Fe$^{3+}$, no static moment modulation can exist at $T=0$ and therefore, the elliptical structure in phase II should progressively transform into a circular structure as $T \to$\,0. In other words, the Fe$^{3+}$ moment on every site should be saturated, i.e. the $y$ ratio should approach 1 as temperature decreases. Our mean $y$ value of 0.93 at 1.46\,K is indeed rather close to 1, but it is definitely lower than 1 since we checked that the 1.46\,K spectrum cannot be correctly fitted with $y$=1. When extrapolating the $y(T)$ thermal variation to zero temperature (see insert of Fig.\ref{hhf}), one obtains $\simeq$0.94, which is lower than 1. Since this is not allowed for Fe$^{3+}$, we interpret this deviation from unity as due to the anisotropy of the magnetic hyperfine interaction itself, which should be independent of temperature. This hyperfine anisotropy is documented for BiFeO$_3$ \cite{zalessky}, and it can be shown that the angular dependence of the modulus of the hyperfine field can be approximated by:
\begin{equation}
H_{hf}(\varphi) = H_{hf}^{//}\ \cos^2 \varphi + H_{hf}^\perp\ \sin^2 \varphi,
\label{anis} 
\end{equation}
where $H_{hf}^{//}$ and $H_{hf}^\perp$ are the main components of the elliptical-like trajectory of {\bf H$_{hf}$}. For values of $y_a=H_{hf}^\perp/H_{hf}^{//}$ close to 1, it is not possible to distinguish the spectral effects of an elliptical dependence due to a moment modulation (expression (\ref{ell})) from those due to the anisotropy of the hyperfine interaction (expression (\ref{anis})). At 1.46\,K, the $y$ value reflects solely the hyperfine anisotropy, but at higher temperature, the dominant contribution to the ratio $y$ is the elliptical moment configuration. Regarding the neutron diffraction derived ratio $y_n \sim$\,0.8 at 2\,K, we have no explanation for such a low value, since it measures directly the ratio of the moment elliptical axes and thus should be much closer to 1.

In another SDW ferroelectric, FeTe$_2$O$_5$Br, an oblate elliptical incommensurate magnetic structure of Fe$^{3+}$ moments has been observed to persist down to 0.053\,K \cite{preg1,preg2}, with a quite small $y$ ratio of 0.37, seemingly violating the ``single-valued moment'' rule for Fe$^{3+}$ as $T \to$\,0. However, a fluctuating disordered moment component has been inferred from $\mu$SR measurements down to very low temperature, which should restore a single ``static'' Fe$^{3+}$ moment on each site as $T \to$\,0. This is confirmed by our M\"ossbauer data in FeTe$_2$O$_5$Br \cite{unpub} at 4.2\,K, which shows a well resolved magnetic hyperfine spectrum with a single hyperfine field of $\simeq$44\,T. This is in good agreement with the moment value of 4\,\mub\ quoted in Ref.\cite{preg1} using the standard hyperfine constant of 11\,T/\mub. The hyperfine Larmor period associated with the $^{57}$Fe magnetic hyperfine interaction is $\tau_M \sim 10^{-8}$\,s, so the fluctuation time of the disordered spin component in FeTe$_2$O$_5$Br must be rather slow, longer than $\tau_M$. Therefore, the persistent spin dynamics at play in FeTe$_2$O$_5$Br down to the lowest temperature does not violate the ``single-valued moment'' rule for Fe$^{3+}$ as $T \to$\,0.

\section{Conclusion}

$^{57}$Fe absorption M\"ossbauer spectra have been recorded in the two incommensurate magnetic phases in FeVO$_4$. They are in very good agreement with the neutron diffraction results in this compound and represent a good illustration of the spectral shapes associated with magnetic phases with incommensurate moment modulations, in the present case planar elliptic and collinear sine-wave. We observe that the ratio of minor to major elliptical hyperfine field axes is not exactly unity as $T \to$\,0, which we interpret as due to the anisotropy of the magnetic hyperfine interaction since no moment modulation can exist for Fe$^{3+}$ as $T \to$\,0. M\"ossbauer spectroscopy is an important complementary method of neutron diffraction for studying Fe (or Sn) containing moment modulated magnetic phases, yielding rather precise values for the characteristics of the magnetic structures and often leading to a more thorough understanding of the system under study.


\begin{thebibliography}{00}

%% \bibitem{label}
%% Text of bibliographic item
\bibitem{smolenski}
G.A. Smolenskii, I.E. Chupis, Sov. Phys. Usp. 25 (1982) 475

\bibitem{mostovoy}
M. Mostovoy, Phys. Rev. Lett. 96 (2006) 067601

\bibitem{tokura}
Y. Tokura, S. Seki, N. Nagaosa, Rep. Prog. Phys. 77 (2014) 076501

\bibitem{smith}
R.T. Smith, G.D. Achenbach, R. Gerson, W.J. James, J. Appl. Phys. 39 (1968) 70

\bibitem{sosnowska}
I. Sosnowska, T. Peterlin-Neumaier, E. Steichele, J. Phys. C 15 (1982) 4835 

\bibitem{kadomtseva}
A. Kadomtseva, A. Zvezdin, Y. Popov, A. Pyatakov, G. Vorob'ev, JETP Letters 79 (2004) 571 

\bibitem{kimura}
T. Kimura, S. Ishihara, H. Shintani, T. Arima, K.T. Takahashi, K. Ishizaka, Y. Tokura, Phys. Rev. B 68 (2003) 060403(R)

\bibitem{goto}
T. Kimura, T. Goto, H. Shintani, K. Ishizaka, T. Arima, Y. Tokura, Nature (London) 426 (2003) 55 

\bibitem{hur}
N. Hur, S. Park, P. A. Sharma, J. S. Ahn, S. Guha, S.W. Cheong, Nature (London) 429 (2004) 392

\bibitem{yoshimori}
A. Yoshimori, J. Phys. Soc. Jpn. 14 (1959) 807

\bibitem{dzialo}
I. Dzyaloshinski, JETP 19 (1964) 960 [Zh. Eksp. Teor. Fiz. 46 (1964) 1420]

\bibitem{daoud}
A. Daoud-Aladine, B. Kundys, C. Martin, P.G. Radaelli, P.J. Brown, C. Simon, L.C. Chapon, Phys. Rev. B 80 (2009) 220402(R)

\bibitem{kundys}
B. Kundys, C. Martin, C. Simon, Phys. Rev. B 80 (2009) 172103

\bibitem{kenzel}

M. Kenzelmann, A.B. Harris, S. Jonas, C. Broholm, J. Schefer, S.B. Kim, C.L. Zhang, S.W. Cheong, O.P. Vajk, J.W. Lynn, Phys. Rev. Lett. 95 (2005) 087206

\bibitem{levinson}
L.M. Levinson, B.M. Wanklyn, Jour. Solid State Chem. 3 (1971) 131

\bibitem{robertson}
B. Robertson, E. Kostiner, Jour. Solid State Chem. 4 (1972) 29

\bibitem{zhang}
J. Zhang, L. Ma, J. Dai, Y. P. Zhang, Zhangzhen He, B. Normand, Weiqiang Yu, Phys. Rev. B 89 (2014) 174412

\bibitem{palewicz}
A. Palewicz, T. Szumiata, R. Przenioslo, I. Sosnowska, I. Margiolaki, Solid State Comm. 140 (2006) 359

\bibitem{lebeugle}
D. Lebeugle, D. Colson, A. Forget, M. Viret, P. Bonville, J.F. Marucco, S. Fusil, Phys. Rev. B 76 (2007) 024116

\bibitem{zalessky}
A.V. Zalessky, A.A Frolov, T.A. Khimich, A.A. Bush, V.S. Pokatilov, A.K. Zvezdin, Europhys. Lett. 50 (2000) 547

\bibitem{marty}
K. Marty, P. Bordet, V. Simonet, M. Loire, R. Ballou, C. Darie, J. Kljun, P. Bonville, O. Isnard, P. Lejay, B. Zawilski, C. Simon, Phys. Rev. B 81 (2010) 054416

\bibitem{deutsch}
M. Deutsch, P. Bonville, A.V. Tsvyashchenko, L.N. Fomicheva, F. Porcher, F. Damay, S. Petit, I. Mirebeau, submitted to Phys. Rev. B
 
\bibitem{preg1} 
M. Pregelj, O. Zaharko, A. Zorko, Z. Kutnjak, P. Jegli\u{c}, P.J. Brown, M. Jagidi\u{c}, Z. Jagli\u{c}i\'c, H. Berger, D. Ar\u{c}on, Phys. Rev. Lett. 103 (2009) 147202

\bibitem{preg2}
M. Pregelj, A. Zorko, O. Zaharko, D. Ar\u{c}on, M. Komelj, A.D. Hiller, H. Berger, Phys. Rev. Lett. 109 (2012) 227202

\bibitem{unpub}
D. Colson, A. Forget, P. Bonville, unpublished results
 
\end{thebibliography}
\end{document}